\documentclass[showpacs,amsmath,amssymb,onecolumn,pra]{revtex4-1}

\usepackage[dvips]{graphicx} 
\usepackage{amsfonts}
\usepackage{amssymb}
\usepackage{amscd}
\usepackage{amsmath}    
\usepackage{enumerate}
\usepackage{epsfig}
\usepackage{subfigure}
\usepackage{xcolor}
\usepackage{amsthm}
\usepackage{framed}
\usepackage{multirow}
\usepackage{color}

\usepackage{hyperref}

\newcommand{\ket}[1]{\mbox{$\left| #1 \right\rangle$}}

\newcommand{\comments}[1]{}
\newcommand{\aver}[1]{\langle #1 \rangle}
\newcommand{\bea}{\begin{eqnarray}}
\newcommand{\eea}{\end{eqnarray}}
\newcommand{\Int}{\mathbb{Z}}

\begin{document}
\preprint{APS/123-QED}

\title{Continuous-variable quantum authentication of physical unclonable keys}
\date{\today}

\author{Georgios M. Nikolopoulos}
\email{nikolg@iesl.forth.gr}

\affiliation{Institute of Electronic Structure \& Laser, 
FORTH, P.O. Box 1385, GR-70013 Heraklion, Greece}

\author{Eleni Diamanti}
\affiliation{Laboratoire d'Informatique de Paris 6, CNRS, UPMC-Sorbonne Universit\'es, 4 place Jussieu, 75005 Paris, France}

\begin{abstract}
We propose a scheme for authentication of physical keys that are materialized by optical multiple-scattering media. The authentication relies on the optical response of the key when probed by randomly selected coherent states of light, and the use of standard wavefront-shaping techniques that direct the scattered photons coherently to a specific target mode at the output. The quadratures of the electromagnetic field of the scattered light at the target mode are analysed using a homodyne detection scheme, and the acceptance or rejection of the key is decided upon the outcomes of the measurements. The proposed scheme can be implemented with current technology and offers collision resistance and robustness against key cloning.
\end{abstract}


\maketitle

\section{Introduction}

Entity authentication (sometimes also referred to as identification) is one of the most important cryptographic tasks, in which one party (the verifier) obtains assurance that the identity of another party (the claimant) is as declared, thereby preventing impersonation \cite{handbook}.
Techniques for identification typically rely on (i) something that the claimant knows (e.g., a secret password or numerical key); (ii) something that the claimant possesses (e.g., a physical token or card); or (iii) something inherent (e.g., biometrics).
The first two techniques are purely cryptographic and are used extensively for everyday tasks (such as transactions in automatic teller machines). High levels of security can be achieved by means of dynamic entity authentication protocols (EAPs) that combine techniques (i) and (ii), through a {\em challenge-response} mechanism \cite{handbook,handbook2}.
More precisely, before any authentication, the user is given a physical key (token or smart card) and a short personal identification number (PIN), which has to be kept secret. The authentication then relies on a publicly-known cryptographic algorithm, such as for instance a symmetric algorithm involving a numerical key that is shared between the verifier and the token. First, the PIN is used to verify the user to the token; if the PIN is correct, the verifier proceeds by generating a number of random and independent numerical challenges, and for each one of them the token computes a response based on the implemented algorithm and the shared key. The user is authenticated only if all of the responses agree with the ones expected by the verifier. An impersonation attack against such a dynamic EAP is difficult but not impossible, especially when the PIN is not well protected. The main weakness of the protocol stems from the fact that traditional physical keys can be cloned \cite{Horstmeyer15}, thereby enabling potential hackers to impersonate successfully legitimate users.

The development of cloning-resistant EAPs is of particular importance for the field of cryptography,  and optical schemes are currently considered to be among the most promising candidates \cite{Horstmeyer15}. In optical EAPs, the physical key is materialized by an optical multiple-scattering random medium, which is probed (or {\em interrogated}) by light pulses ({\em probes}) \cite{Horstmeyer15,Pappu02,PappuPhD,Buchetal05,Tuyls05,Skoric08,Skoric13,Horstmayer13,Goorden14,Iesl16a,Iesl16b}.
Such disordered keys are considered to be practically unclonable because their full characterization involves a large number of degrees of freedom, and they are usually referred to as physical unclonable keys (PUKs) or functions (PUFs). Their optical {\em response} to a probe depends on the details of their internal disorder, as well as on different parameters of the probe. Typically, an optical EAP has two stages \cite{Pappu02,PappuPhD}. The {\em enrolment stage} takes place  only once, before the key is given to the user, and aims at its full characterization by the authority responsible for the distribution of the keys. To this end, the key is subject to a large number of random {\em challenges} (i.e., it is interrogated by large number of probes with different parameters), and all of the challenge-response pairs (CRPs) are stored in a database together with the PIN. In the {\em verification stage}, the user inserts his key in a verification device and types in his secret PIN. If the PIN is correct, the verifier has to decide whether the key with the given PIN is authentic or not. Assuming that the verifier has access to the database, the verification is achieved by interrogating the key with a moderate number of probes, whose parameters are chosen at random from the recorded challenges in the  database, and by checking whether the corresponding responses agree with those in the database.

The cloning resistance of optical PUKs is sufficient for preventing impersonation attacks in a tamper-resistant scenario, typically discussed and analysed in the literature \cite{Horstmeyer15,Pappu02,PappuPhD,Iesl16a}, where an adversary does not have access to the probes. In scenarios where an adversary may actually have access to the probes that are sent to the optical PUK, the nature of these probes (challenges) plays a significant role in the security of the authentication protocol. When the probe is classical light, the controlled parameters are classical quantities such as the incidence angle, the intensity, or the wavefront of the field \cite{Pappu02,PappuPhD,Buchetal05,Horstmayer13}. Hence, an adversary who has access to the verification set-up is, in principle, able to read out, copy, and manipulate the classical information carried by the probes, without being detected.
The security of optical EAPs may increase considerably by using quantum instead of classical probes. In this case, information gain about the quantum state of a probe is limited by fundamental laws of quantum physics, and can be obtained only at the cost of  disturbing the quantum state of the probe \cite{book}.
In this spirit,  Goorden {\em et al}  proposed and implemented an EAP, in which the challenges are encoded on attenuated laser pulses with shaped wavefronts \cite{Skoric13,Goorden14}. The implementation of this scheme requires photon-counting detectors, and acceptance or rejection of a key is decided upon the number of photodetection events.

Here we propose a new optical EAP, in which information is carried by the continuous quadrature components of the quantized electromagnetic field of the probe. Such a continuous-variable encoding has been shown to offer practical advantages in quantum key distribution \cite{Jouguet13}. The implementation of our protocol relies on standard wavefront-shaping and homodyne-detection techniques, and is within reach of current technology. Assuming a tamper-resistant verification set-up, we show that the protocol offers highly desirable features, such as collision resistance and robustness against key cloning, which are necessary for the protocol to be useful in practice \cite{Pappu02,PappuPhD}.

\begin{figure}[t]
\centering
\includegraphics[width=0.8\linewidth,angle=0]{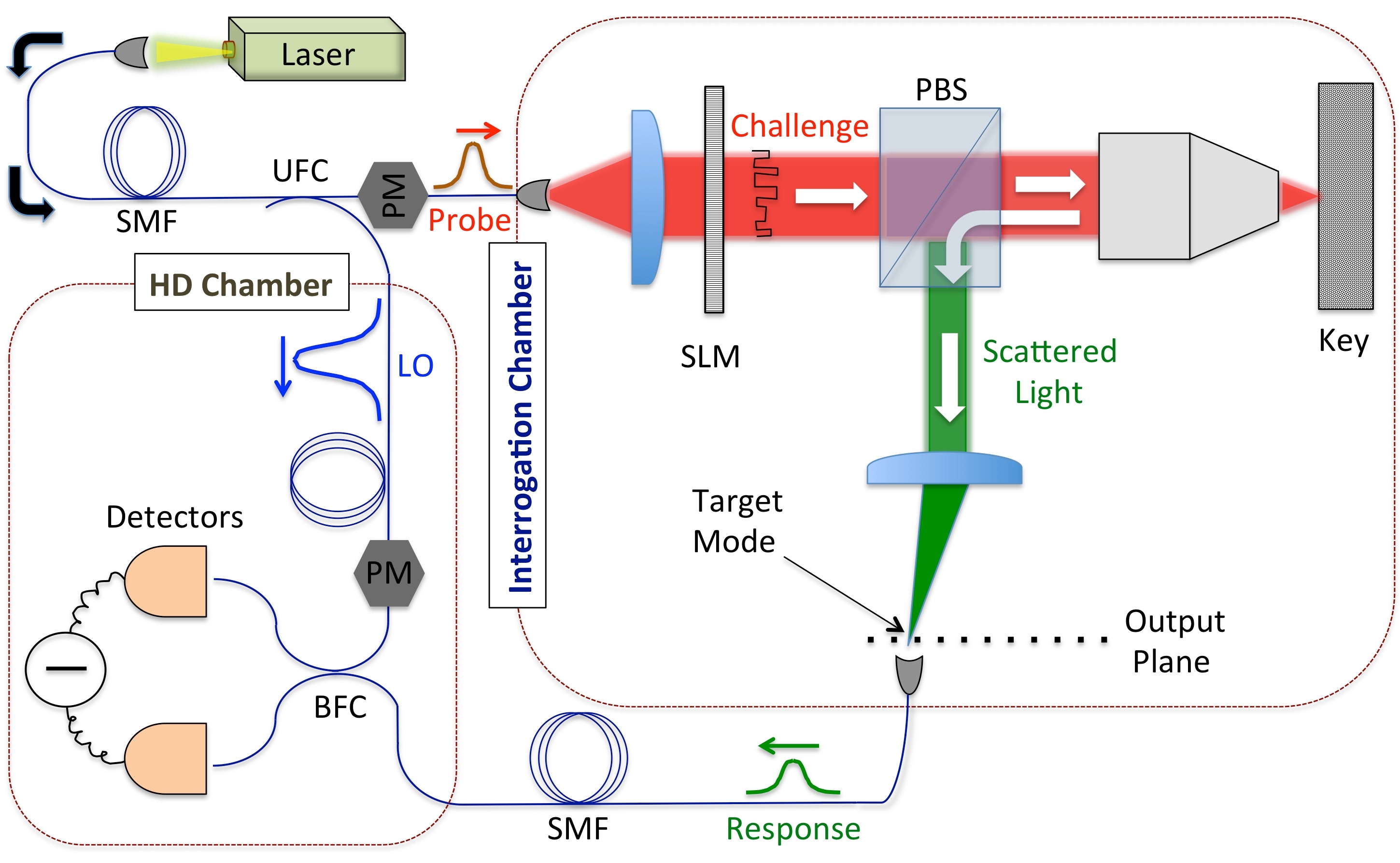}
\caption{Schematic representation of the authentication protocol. The output of the laser is injected into a single-mode fiber (SMF) and then split, using an unbalanced fiber coupler (UFC), into a large fraction that serves as the local oscillator (LO) and a small fraction that serves as the probe in the verification. The phase of the probe relative to the LO is adjusted using a phase modulator (PM), and the challenge is obtained by modulating the wavefront of the probe using a phase-only spatial-light modulator (SLM). The challenge is then focused on the key, and the scattered (reflected) light is coupled out by means of a polarizing beam splitter (PBS), which ensures the collection of light that has undergone multiple scattering in the key \cite{Defienne14}. The phase mask of the SLM is adjusted so that the scattered light is focused on one of the transverse modes of the output plane, where it is coupled to a SMF. The quadratures of the electric field of the scattered light are measured using a standard homodyne detection (HD) set-up involving a phase modulator in the LO path, a balanced fiber coupler (BFC) and two photodiodes. 
The laser source, the interrogation chamber and the HD chamber are considered to be well-separated and connected via SMFs.
}
\label{fig1}
\end{figure}


\section{Results}


\subsection{Authentication set-up}

A realization of the proposed EAP is shown in Fig.~\ref{fig1}, and consists of the probe state preparation set-up, the interrogation chamber, and the homodyne-detection (HD) 
set-up (chamber). Except for the HD, the scheme is similar to the wavefront-shaping set-up used for the control of light scattered by a disordered multiple-scatering medium 
(to be referred to hereafter as the key)
\cite{Vellekoop07,Vellekoop08,Mosk12,Huisman14,Huisman15,Poppoff10,Poppoff11,Defienne14,Huisman14b,Wolterink16}.
The laser beam at wavelength $\lambda$ is split into two parts: a weak probe that is sent to the wavefront shaping set-up, and a strong local oscillator, which will serve as a reference in the HD of the scattered light. The key is assumed to have a slab geometry with thickness $L$ and mean free path $l\ll L$.  In the {\em diffusive regime}, i.e. for $\lambda\ll l\ll L\ll L_{\rm abs}$, where $L_{\rm abs}$ is the absorption length, light undergoes multiple scattering events in the key, and the process can be described in terms of a finite number of discrete input and output transverse spatial modes \cite{Goodman1,book2,Lodahl_prl05,Lodahl_oe06}. 
Using a phase-only spatial light modulator (SLM), one can control the phases of the incoming modes, thereby directing coherently the main part of the scattered light into a prescribed outgoing mode (to be referred to hereafter as the target mode) \cite{Vellekoop07,Vellekoop08,Mosk12,Huisman14,Huisman15,Poppoff10,Poppoff11,Defienne14,Huisman14b,Wolterink16}.   For a given key, one can select different target modes by changing accordingly the phase mask of the SLM. Moreover, different output transverse modes can be addressed by a single-mode fiber (SMF), which can be translated on the output (optical) plane in a controlled manner, provided that the overall imaging system is optimized so that the diameter of a single speckle grain matches the diameter of the mode of the SMF \cite{Defienne14,Huisman14b}.

\begin{figure}[t]
\centering
\includegraphics[width=0.5\linewidth,angle=0]{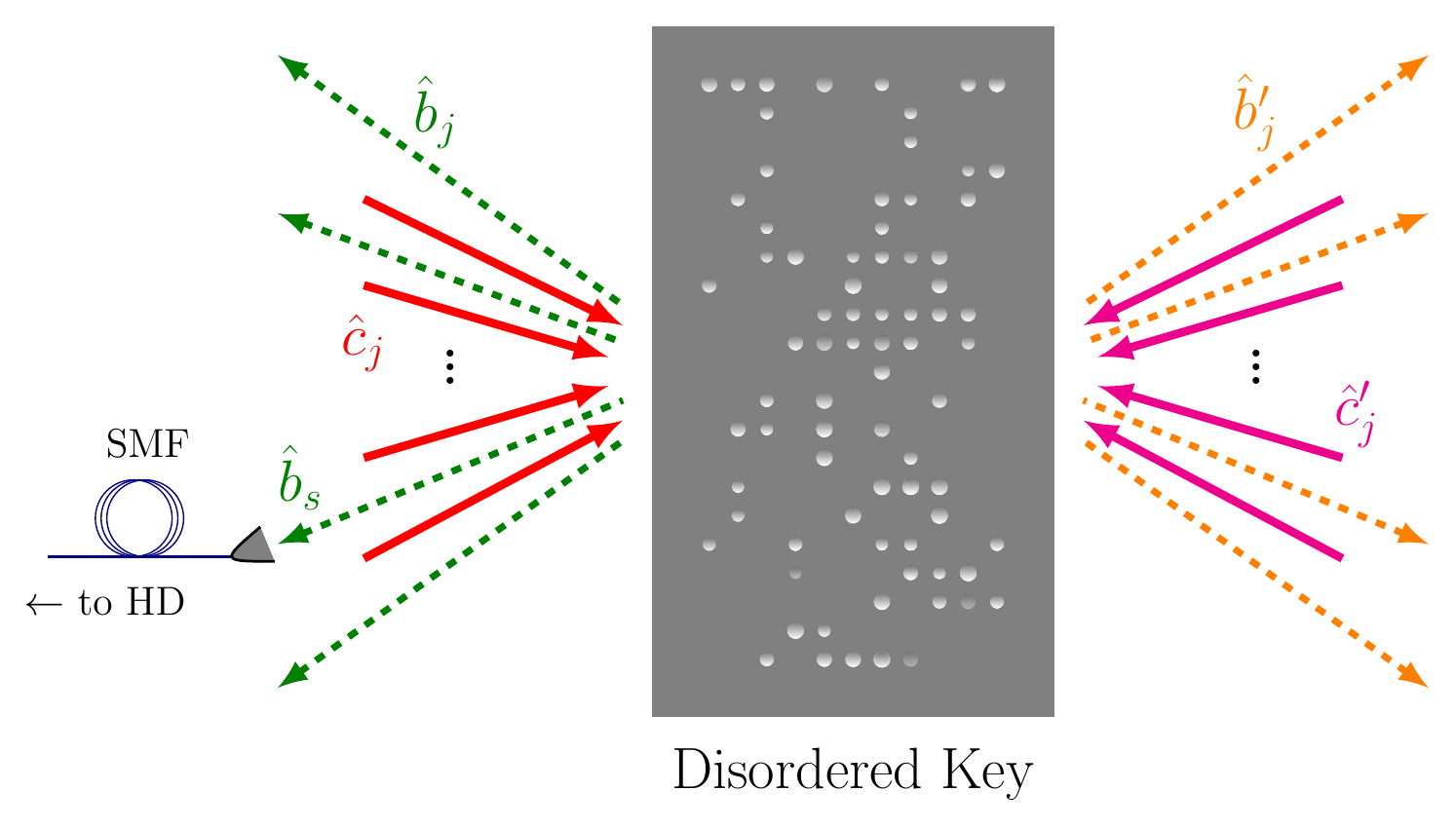}
\caption{Schematic representation of the incoming (solid arrows) and outgoing (dashed arrows) modes with respect to the disordered  key. All of the ${\cal N}$ incoming modes on the right of the key (magenta arrows) are initially in vacuum, which implies $\aver{\hat{c}_j^\prime}=0$ for all $j=1,2,\ldots, {\cal N}$. Only incoming modes on the left of the key (red solid arrows) are initially excited, with their relative phases  optimized so that the main part of the reflected light is collected in a particular outgoing mode, which is addressed by a SMF (corresponding annihilation operator $\hat{b}_s$). The collected light is transferred to the HD chamber, where it is analysed. The transmitted light (dashed orange arrows on the right of the key) is not monitored in our set-up, and the corresponding equations are not of relevance. 
}
\label{fig1b}
\end{figure}


\subsection{Formalism}

Throughout this work we  adopt the Heisenberg picture, because it facilitates the comparison with the classical setting. Following existing literature \cite{Goodman1,book2,Lodahl_prl05,Lodahl_oe06},  
for the sake of simplicity we assume ${\cal N}$ incoming and  ${\cal N}$ outgoing modes on each side of the key  (see Fig. \ref{fig1b}). However, our analysis is expected to remain valid more generally, with appropriate adjustment of the formulas below. In the set-up of Fig. \ref{fig1} the target mode is one of the ${\cal N}$ outgoing modes on the left of the key 
(labelled by $s\in \{1,2,\ldots,{\cal N}\}$ ), which is coupled to a SMF, and let $\hat{b}_{s}$ be the corresponding annihilation operator for a photon (see Fig. \ref{fig1b}). Only incoming modes on the left of the key  are initially populated and are controlled by the SLM, whereas all the incoming modes on the right of the key are in vacuum. Hence, one  readily obtains \cite{Lodahl_prl05,Lodahl_oe06}
\bea
\label{heisenberg:eom1}
\aver{\hat{b}_{s}} &=& \sum_{j=1}^{\cal N} R_{s,j} \aver{\hat{c}_j},
\\
\aver{\hat{b}_{s}^\dag \hat{b}_{s}} &=& \sum_{j,j^\prime=1}^{\cal N}  R_{s,j^\prime}^\star R_{s,j} \aver{\hat{c}_{j^\prime}^\dag\hat{c}_j},
\label{heisenberg:eom2}
\eea
where $\aver{\cdot}$ denotes quantum mechanical expectation value, $\hat{c}_j$ is the annihilation operator for a photon in the  $j-$th  incoming mode on the left of the key, and $\{R_{s,j}\}$ are the electric-field reflection coefficients from the $j-$th incoming  mode to the target mode. The latter depend on the realization of the disorder in the medium and can be treated as independent complex Gaussian random variables  that satisfy  \cite{Goodman1,book2,Lodahl_prl05,Lodahl_oe06}
\bea
&&\overline{R_{m,j}} = 0, \quad
\overline{|{R}_{m,j}|^2} = {\cal N}^{-1}\left (1-l/L\right ):={\cal V},
\label{Rstat:eq}
\eea
where the over-line denotes (classical) ensemble average over all disorder realizations. 
The main assumptions underlying this ``Gaussian-statistics model"  are summarized in the Methods.

By analogy with Eqs. (\ref{heisenberg:eom1}) and  (\ref{heisenberg:eom2}), the coupling of the SMF at the input of the verification set-up to the incoming modes at the exit of the SLM can be modelled by equations of the form
\bea
\label{heisenberg:eom3}
\aver{\hat{c}_{j}} &=& g_j \aver{\hat{a}} e^{{\rm i}\phi_j},
\\
\aver{\hat{c}_{j^\prime}^\dag \hat{c}_j} &=& g_{j^\prime}^\star g_j \aver{\hat{a}^\dag \hat{a}} e^{{\rm i}(\phi_j-\phi_{j^\prime})}, 
\label{heisenberg:eom4}
\eea
where $\hat{a}$ is the annihilation operator for a photon in the mode of the fiber and $g_j$ is the electric-field transmission coefficient from the fiber's mode to the $j-$th incoming mode. Analogous models have been employed in various contexts in physics 
for the description of outcoupling from cavities and waveguides \cite{book3,Yariv}.
The specific form of the coefficients $\{g_j\}$ depends on  the details of the mechanism that governs the coupling between the mode of the fiber and the modes at the exit of the SLM, and is not needed for the purpose of this work. For what follows, however, it is important to emphasize that these coefficients are in general complex numbers that satisfy $\sum_{j} |g_j|^2=\tau\leq 1$, where the constant $\tau$ accounts for possible losses. Contrary to the reflection coefficients $\{R_{s,j}\}$, the coefficients $\{g_j\}$ are independent of the key, and are expected to be fully determined by the details of the verification set-up (e.g., wavelength of the light, cross section of the fiber, separation of various elements, etc). 
Throughout this work we will be interested in a fixed  verification set-up, with publicly known specifications, which means that $\{g_j\}$ have to be considered as publicly known constants as well.

The set of angles $\Phi^{(s)}:=\{\phi_{1},\ldots,\phi_{{\cal N}}\}$ in 
Eq. (\ref{heisenberg:eom3})  refer to the phase mask of the SLM, and may or may not be 
optimized with respect to the particular target mode (denoted by $s$). In the 
absence of optimization, the scattered light is distributed among the various modes at the  
output, with the precise form of the corresponding intensity distribution (speckle pattern) 
depending on the realization of disorder. By optimizing 
the phase mask of the SLM one can maximize the concentration of scattered light in  
the target mode $s$. The optimization may involve feedback algorithms, 
in which the phase mask of the SLM is optimized with respect to the intensity (or power) of the 
scattered light in the target mode\cite{Vellekoop08,Yilmaz13,Anderson14}. Alternatively, 
an optimal phase mask can be found by means of the experimental estimation of 
the scattering matrix of the key \cite{Poppoff11}.  
The directional concentration of scattered light in the target mode is never complete, because light will be unavoidably scattered in other outgoing modes as well. Hence, the amount of control one has over the propagation of light in the disordered key is usually quantified by the intensity enhancement ${\cal E}$ i.e., the ratio of the intensity in the target mode after optimization, to the ensemble-average intensity in the absence of optimization \cite{Vellekoop08,Yilmaz13,Anderson14}.
Generalizing this classical definition to a quantum setting we have
\bea
{\cal E} = \frac{\aver{\hat{b}_s^\dag\hat{b}_s}_{\rm o}}{\overline{\aver{\hat{b}_s^\dag\hat{b}_s} }_{\rm no}},
\label{E:def}
\eea
where $\aver{\hat{b}_s^\dag\hat{b}_s}_{\rm o}$ is the mean number of scattered photons in the target mode in the presence of an optimized SLM  for a single realization of disorder, whereas $\overline{\aver{\hat{b}_s^\dag\hat{b}_s}_{\rm no} }$ in the denominator is the 
corresponding ensemble-average mean number of photons in absence of optimization. From now on, $\Phi_{\rm opt}^{(s)}$ will denote the  optimal phase mask that maximizes the number of scattered photons in the target mode $s$, for a given key. For the sake of simplicity, the dependence of $\Phi_{\rm opt}^{(s)}$ on the key (i.e., on the realization of the disorder), will not be explicitly shown. 

The above formalism is rather general, in the sense that so far there have been no explicit assumptions about the quantum state of the probes that are used in the interrogation of the key. The proposed EAP uses coherent states of light, and relies on standard HD techniques. In particular, we treat the states of the local oscillator (LO) and the probe as single-mode coherent states, $\ket{\alpha_{LO}}$ and $\ket{\alpha_P} = \ket{\sqrt{\mu_P}e^{{\rm i}\varphi_P}}$ respectively, where $\mu_P$ is the mean number of photons in the probe and $\varphi_P$ is a relative phase with respect to the LO.
The coherent state $\ket{\alpha_P} $ is an eigenstate of  $\hat{a}$ with eigenvalue $\alpha_P$, 
and thus 
\bea
\aver{\hat{a}} = \alpha_P,\quad\aver{\hat{a}^\dag} = \alpha_P^\star,\quad{\rm and}\quad \aver{\hat{a}^\dag \hat{a}} = |\alpha_P|^2.
\label{a:ev}
\eea 
Using Eqs. (\ref{heisenberg:eom1}), (\ref{heisenberg:eom3}) and (\ref{a:ev}) we obtain
\bea
\aver{\hat{b}_{s}} = \left [ \sum_{j=1}^{\cal N} R_{s,j} g_j e^{{\rm i}\phi_j} \right ]\alpha_P, 
\label{avBnonuni}
\eea
with the case of uniform illumination of the SLM obtained for $|g_j|=\sqrt{\tau/{\cal N}}$.
The analogy of Eq. (\ref{avBnonuni}) to equations used in the analysis and the implementation of wavefront-shaping with classical light sources \cite{Vellekoop08}, stems from the use of coherent probe states, and the preservation of coherence throughout the wavefront-shaping and the scattering. The latter is reflected in the linearity of the input-output equations   (\ref{heisenberg:eom1})-(\ref{heisenberg:eom4}), which in view of Eqs. (\ref{a:ev}) imply 
\bea
\aver{\hat{b}_{s}^\dag\hat{b}_s}=|\aver{\hat{b}_{s}}|^2.
\label{b:decor}
\eea

Equation (\ref{avBnonuni}) determines the expectation value of the electric field in 
the mode of the fiber. The quadrature amplitudes of the field can be measured by means 
of  HD, with the LO serving as the required reference \cite{book3}. 
By adjusting the LO phase $\theta$, one measures the generalized quadrature amplitude 
$\hat{Q}_s(\theta) = (\hat{b}_s^\dag e^{{\rm i}\theta}+\hat{b}_s e^{-{\rm i}\theta})/\sqrt{2}$.
Assuming that the LO field is much stronger than the total scattered field 
(i.e., for $|\alpha_{LO}|\gg|\alpha_P|$),  the outcome of such a measurement is a real random number 
$q$ which, to a good 
accuracy, follows a Gaussian distribution \cite{Raymer95}
\bea
{\rm Pr}(q|\aver{\hat{Q}_s(\theta)}) =  \frac{1}{\sqrt{2\pi \sigma^2}}
\exp\left \{ -\frac{[q- \aver{\hat{Q}_s(\theta)}]^2}{2\sigma^2 } \right \},
\label{PrHD:eq}
\eea
with the shot noise $\sigma \simeq 1/\sqrt{2\eta}$, where  $\eta\leq 1$ is  the detection efficiency. 
Hence, the measurement of the quadrature $\hat{Q}_s(\theta)$ is equivalent to sampling from the distribution (\ref{PrHD:eq}).
Throughout this work we focus on the measurement of the real $\hat{X}_s$ and imaginary $\hat{Y}_s$  quadratures, corresponding to $ \hat{Q}_s(0)$ and $\hat{Q}_s(\pi/2)$, respectively. The corresponding Gaussian photocount distributions are centred at $\aver{\hat{X}_{s}}=\sqrt{2}{\rm Re}(\aver{\hat{b}_{s}})$ and $\aver{\hat{Y}_{s}}=\sqrt{2}{\rm Im}(\aver{\hat{b}_{s}})$, for $\theta=0$ and $\pi/2$, respectively.
In the framework of our protocol, we will refer to
${\cal R}_s=\aver{\hat{X}_{s}}+{\rm i}\aver{\hat{Y}_{s}}=|{\cal R}_s|e^{{\rm i}\varphi_s} $ as the response of the key to the probe state $\ket{\alpha_P}$.

The above expressions and observations are applicable to the cases of both optimized and non-optimized SLM.
Let ${\cal R}_s^{({\rm no})} := \aver{\hat{X}_{s}}_{\rm no}+{\rm i}\aver{\hat{Y}_{s}}_{\rm no} = \sqrt{2}\aver{\hat{b}_{s}}_{\rm no} $
denote the response of the key in the absence of SLM optimization, where $\aver{\hat{X}_{s}}_{\rm no}$ and
$\aver{\hat{Y}_{s}}_{\rm no}$ are the centres of the photocount distributions for
$\theta=0$ and $\pi/2$, respectively.
Both of $\aver{\hat{X}_{s}}_{\rm no}$ and $\aver{\hat{Y}_{s}}_{\rm no}$ depend on the 
realization of the disorder, and statements can be made only for ensemble averages.
Using Eqs. (\ref{avBnonuni}) and (\ref{b:decor}) for 
$\aver{\hat{b}_{s}} =\aver{\hat{b}_{s}}_{\rm no} $, as well as  the independence of $\{R_{s,j}\}$  and 
Eqs. (\ref{Rstat:eq}), one readily obtains
\bea
\overline{\aver{\hat{b}_{s}}_{\rm no} } = 0,\quad 
\overline{\aver{\hat{b}_{s}^\dag \hat{b}_s}_{\rm no} } = {\cal V}\mu_{\rm c}, 
\label{b:nonopt}
\eea
where $\mu_{\rm c}:=\tau\mu_P$ is the total mean number of photons in the challenge, at the exit of the SLM.
Hence, using the above relation between   ${\cal R}_s^{({\rm no})} $ and $\aver{\hat{b}_{s}}_{\rm no} $ we have 
\bea
\overline{\aver{\hat{X}_{s}}_{\rm no} }=\overline{\aver{\hat{Y}_{s}}_{\rm no} }=0,\quad \overline{(\aver{\hat{X}_{s}}_{\rm no})^2} + \overline{(\aver{\hat{Y}_{s}}_{\rm no})^2} = 2{\cal V}\mu_{\rm c}.
\label{XY:nonopt}
\eea
For a given realization of disorder, when the SLM is optimized so that the scattered light is mainly 
directed to the target mode $s$, the response of the key will be denoted by 
${\cal R}_s^{({\rm o})} := \aver{\hat{X}_{s}}_{\rm o}+{\rm i}\aver{\hat{Y}_{s}}_{\rm o} = \sqrt{2}\aver{\hat{b}_{s}}_{\rm o} $, and the 
photocount distributions for $\theta=0$ and $\pi/2$ are expected to be centred at 
$\aver{\hat{X}_{s}}_{\rm o}$ and $\aver{\hat{Y}_{s}}_{\rm o}$, respectively. 
Using Eq. (\ref{b:decor})  for $\aver{\hat{b}_{s}} =\aver{\hat{b}_{s}}_{\rm o} $, 
and Eq. (\ref{b:nonopt}) one readily obtains from Eq.  (\ref{E:def}) 
\bea
(\aver{\hat{X}_{s}}_{\rm o})^2+ (\aver{\hat{Y}_{s}}_{\rm o})^2= 2{\cal E} {\cal V}\mu_{\rm c}.
\label{optXY:condition}
\eea

Finally, an important quantity for what follows is the conditional probability for the outcome $q$ in a HD along $\theta$ to fall within the interval (bin)
\[{\mathfrak B}[{\cal R}_s,\theta]=\left [~ |{\cal R}_s|\cos(\varphi_s-\theta)-\delta/2,
|{\cal R}_s|\cos(\varphi_s-\theta)+\delta/2~\right ]\]
for some $\delta$ such that $2\sigma\lesssim \delta<4\sigma$. From Eq. (\ref{PrHD:eq}) and the above discussion, we have
\bea
\label{Pexp:eq}
{\rm Pr}({\rm in}|{\cal R}_s,\theta) =
 {\rm Erf}\left (\frac{\delta}{2\sqrt{2}\sigma}
\right ),
\eea
which is independent of $\theta\in\{0,\pi/2\}$, and depends only on the ratio $\delta/\sigma$. This is because for both values of $\theta$, the  bin is centred at the centre of the Gaussian distribution of Eq.~(\ref{PrHD:eq}). Moreover, Eq. (\ref{Pexp:eq}) is valid for both optimized and non-optimized SLM, provided that the bin is defined for ${\cal R}_s^{({\rm o})}$ and ${\cal R}_s^{({\rm no})}$, respectively. In either case, it should be kept in mind that according to Eq. (\ref{avBnonuni}), ${\cal R}_s^{({\rm o})}$ and ${\cal R}_s^{({\rm no})}$ depend on the probe state $\ket{\alpha_P}$, on the key (through the reflection coefficients of the scattering matrix), as well as on the phase-mask of the SLM. One cannot know ${\cal R}_s^{({\rm o})}$ or ${\cal R}_s^{({\rm no})}$ without knowing all of these pertinent quantities.


\subsection{Entity authentication protocol}

We assume that the set-ups used for the enrolment and the verification stages of the 
EAP are the same. All of their specifications (i.e., losses, imperfections, detection 
efficiency, wavelength of light, etc) together with $\delta$ and convergence parameters 
$\varepsilon,\zeta\ll 1$, are publicly known. 
As will become clear below, $\varepsilon$ and $1-\zeta$ are the error and the confidence 
levels in the verification stage of the EAP, respectively. 
Hence, ${\rm Pr}({\rm in}|{\cal R}_s,\theta)$ becomes a publicly known constant that will 
be denoted by $P_{\rm in}$. 
For the sake of clarity, we will discuss the protocol in the framework of  coherent states 
with the same amplitude but different phases. However, the protocol can also be implemented 
with states that differ both in phase and in amplitude, and the generalization of the following 
results and observations to this case is straightforward.
Let  
\bea
{\mathbb A} =\left \{\ket{\alpha_k}=\ket{\sqrt{\mu_P}e^{{\rm i}\varphi_k}}~:~\varphi_k = 2\pi k/N,\,k \in \Int_{N}\right \}
\label{Aset:def}
\eea
be a publicly known set of coherent probe states, with $\Int_{N} \equiv \{0,1,2,\ldots,N-1\}$ and $N>2$. 
Note that the states in ${\mathbb A} $ are uniquely identified by the values of the integer $k$. 

\subsubsection{Enrolment stage}

Each key is associated with a single target mode ${s}$ chosen at random from the set of  all accessible 
 target modes in the set-up.
In the enrolment stage, the first task of the enroller is to find the optimal phase mask 
$\Phi_{\rm opt}^{(s)}$ for the SLM that directs the scattered light to the particular target mode. 
A classical light source and known techniques \cite{Vellekoop08,Poppoff10,Poppoff11} can be used to this end, because an optimal phase mask works in the same way in the classical and the quantum regimes 
\cite{Defienne14,Huisman14,Wolterink16}. Subsequently, for each $\ket{\alpha_k}\in {\mathbb A} $ the 
key is interrogated by many probes (each one prepared in $\ket{\alpha_k}$), with the phase mask of the 
SLM set to  $\Phi_{\rm opt}^{(s)}$, and for each probe one of the quadratures of the field in the target 
mode is measured. In a standard HD set-up, the enroller has to switch randomly between $\theta=0$ and 
$\theta=\pi/2$ so as to obtain sufficiently large samples for a reliable estimation of both 
$\aver{\hat{X}_s}_{\rm o}$ and  $\aver{\hat{Y}_s}_{\rm o}$, and thus of the corresponding optimized 
response ${\cal R}_s^{({\rm o})}(\alpha_k)$. For the sake of clarity, the dependence of 
${\cal R}_s^{({\rm o})}$ on the scattering matrix of the key and the phase mask of the 
SLM is not 
explicitly shown here. It is essential for each one of the possible probe states, to estimate 
the response ${\cal R}_s^{({\rm o})}(\alpha_k)$ with accuracy higher than the accuracy to be 
used in the verification. For a fixed probe state $\ket{\alpha_k}$, the samples that are 
obtained for the estimation of either of the two quadratures are assumed to be independent 
and identical. Hence, from the central-limit theorem we have that,  with high probability, 
the absolute error in 
the  estimation of either  $\aver{\hat{X}_s}_{\rm o}$ or  $\aver{\hat{Y}_s}_{\rm o}$ does 
not exceed $\xi:=5/\sqrt{M_e}$, where $M_e$ is the sample size used 
in the estimation of one of the quadratures for the given probe state (see Methods below).  
Given that the enrolment is performed only once by the authority that creates and 
distributes the keys,  well before they are given to the users, it is reasonable to assume that the enroller has 
all the freedom to obtain as large samples as needed for the error to satisfy 
$\xi\ll \varepsilon$. Repeating the same procedure for both quadratures and for all of 
the states in ${\mathbb A}$, the enroller can form a list of challenge-response pairs (CRPs), 
with each pair given by  $\{s, k, \Phi_{\rm opt}^{(s)}~|~{\cal R}_s^{({\rm o})}(\alpha_k)\}$, 
which has to be stored in a secure database and will be used for the authentication of the key.

\subsubsection{Verification stage}

When a user gives the key for authentication, the verifier contacts the database over a secure 
authenticated classical channel to obtain the pertinent list of CRPs.  
The verification stage involves $M\gg 1$ identical  sessions, and proceeds as follows.
\begin{enumerate}
\item Set the phase mask of the SLM to $\Phi_{\rm opt}^{(s)}$, and position the SMF at the 
output to match the corresponding target mode $s$.
\item Prepare the probe in  the coherent state $\ket{\alpha_P}$, chosen at random from a 
uniform distribution over ${\mathbb A}$, and send it to the wavefront-shaping set-up.
\item Measure at random the real or the imaginary quadrature of the scattered field in the 
target mode by setting the LO phase to $\theta=0$ or $\pi/2$, respectively. Both quadratures 
are equally probable.
\item Check whether the outcome of the measurement falls within the bin 
${\mathfrak B}[{\cal R}_s^{({\rm o})}(\alpha_P),\theta]$ or not, where $\theta$ is the angle 
that has been chosen in step (3).
\item Repeat steps (2) - (4) $M$ times, and estimate $p_{\rm in}:= M_{\rm in}/M$, 
where $M_{\rm in}$ is the total number of outcomes that have fallen within the bins. 
\item If $|p_{\rm in} - P_{\rm in}|<\varepsilon$ accept the key, otherwise reject.
\end{enumerate}

\begin{figure}[tb]
\centering
\includegraphics[width=0.9\linewidth]{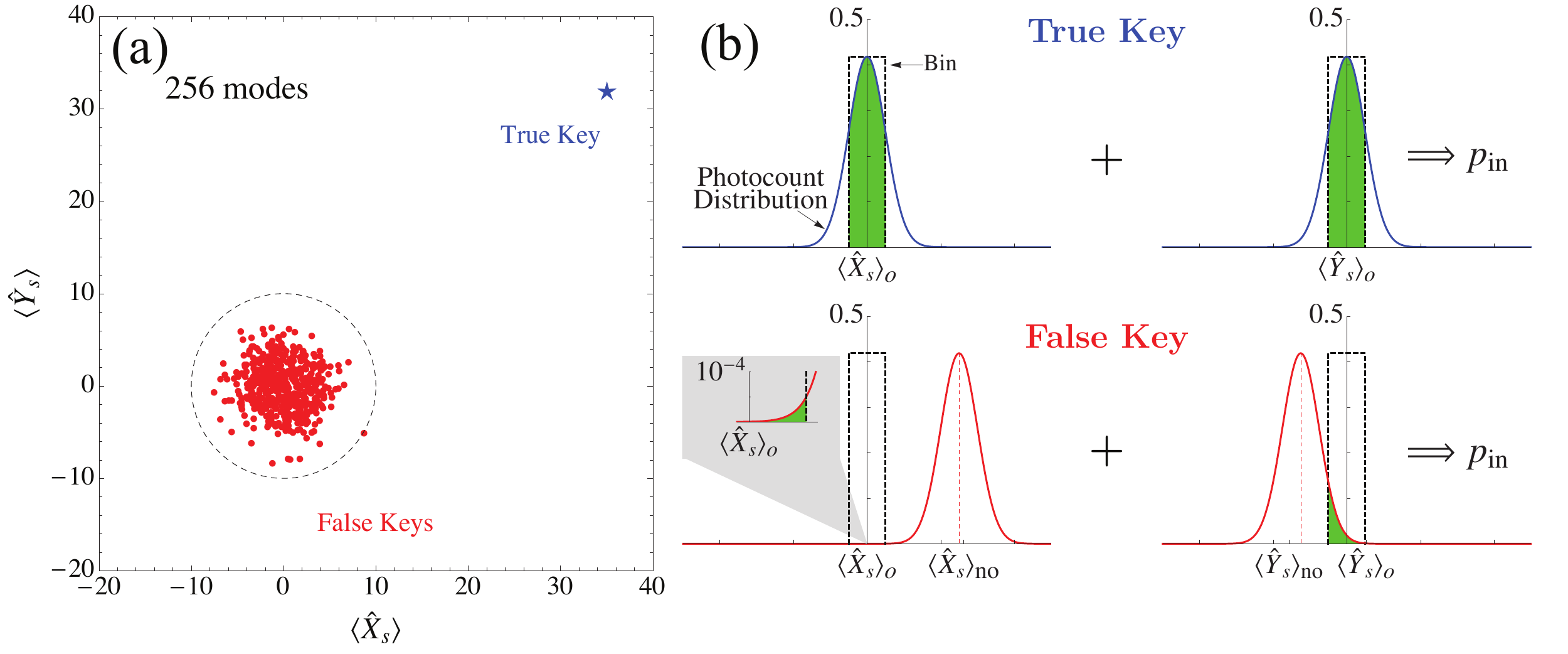}
\caption{(a) Phase-space representation of the typical response of the true key and a false key 
in a single session of the protocol. The red dots refer to 500 randomly chosen false keys, 
while the dashed circle is centred at $(0,0)$ and has radius $\rho_f$ (see text for definition). 
Parameters: ${\cal N}=256$, $\mu_P =  2500$, $\varphi_P=0$, uniform illumination of SLM,  
$\tau=0.8$, $l/L=0.2$. (b) Schematic representation of the estimation of $P_{\rm in}$ in the 
protocol, for the true key and a false key. An estimate $p_{\rm in}$ is obtained by measurements 
of both quadratures. For the true key, the bin is always centred at the center of the photocount 
distribution, whereas for a false key there will always be a shift of the bin relative to the 
photocount distributions, introducing errors in the estimation of $P_{\rm in}$.
}
\label{fig2}
\end{figure}

Given that the verifier is the one who chooses randomly the CRP in each session, he is also able 
to choose the bin so that its centre coincides with the centre of the expected photodetection 
distribution for the true key (namely, ${\rm Pr}(q|\aver{\hat{X}_s}_{\rm o})$ and  
${\rm Pr}(q|\aver{\hat{Y}_s}_{\rm o})$ for $\theta = 0$ and  $\pi/2$, respectively). 
Hence, in the limit of $M\to \infty$, one expects  $p_{\rm in} \to P_{\rm in}$. 
On the contrary, as will be explained below, a false key will result in estimates that deviate 
from $P_{\rm in}$, and thus the verifier could always detect such a key if he could perform 
an arbitrarily large number of sessions. This is, however, not possible in practice. 
Our EAP can be useful in practice only if the verification stage is quick, which means that 
only a moderate number of sessions can be applied during verification. As a result, there will be 
statistical deviations of the empirical probability $p_{\rm in}$ from the theoretical probability 
$P_{\rm in}$, in addition to the deviations that are due to a false key. Distinguishing between 
deviations of different origin is impossible, but the verifier can bound the statistical deviations 
by choosing $M$ sufficiently large. According to the Chernoff bound 
(see Methods) \cite{NikBro16,book4,book5}, when the true key is interrogated 
by $M> M_{\rm th}$ probes, where
\bea
M_{\rm th} :=\frac{3 \ln(2\zeta^{-1})}{\varepsilon^2}
\label{M_th:eq}
\eea
for some $\zeta\ll 1$ and  $\varepsilon \ll P_{\rm in}$, then the probability for the estimate  
$p_{\rm in}$ to deviate from $P_{\rm in}$ by more than $\varepsilon$ is bounded from 
above by $\zeta$, i.e., ${\rm Pr}(|p_{\rm in} - P_{\rm in}|\geq \varepsilon)<\zeta$. 
Hence, for any $M>M_{\rm th}$, the verifier can be $100(1-\zeta)\%$ confident that for the 
true key the statistical deviations cannot exceed $\varepsilon$.
This implies that if  the verifier  obtains an estimate such that $|p_{\rm in} - P_{\rm in}|\geq \varepsilon$, 
then he can be confident that with high probability the observed deviations are due to a false  key.

In closing, we would like to emphasize once more the fundamental difference between the
enrolment and the verification stages. By definition, the enrolment is performed only once, 
by the authority that creates and distributes the keys, and it aims at the accurate 
characterization of a key with respect to its response to all of the possible probe states. 
It is natural, therefore to assume that the enroller has all the time needed so that the 
accuracy in the estimation of the response of the key to a particular probe state, is 
considerably higher than the accuracy in the verification stage.  
By contrast, the verification stage takes place each time the holder of a key has to be 
authenticated, and the verifier has to decide on the acceptance or rejection of a key as 
quickly as possible. Hence, it is crucial for the sample size in the  verification stage to 
be ``small" enough so that it can be obtained within a reasonable period of time 
(say seconds), and at the same time ``large" enough to ensure a reliable verification. 
This point will be made clearer in the following sections.


\subsection{Security aspects}
In order for our EAP to be useful in practice, it has to offer collision resistance and high sensitivity to the randomness of the key \cite{Pappu02,PappuPhD}. Assuming a tamper-resistant verification set-up, in this section we address both of these issues. For the security analysis, it is worth keeping in mind two aspects of the EAP: the phase mask of the SLM is optimized with respect to the true key and a randomly chosen output mode, and in each session of the verification stage the CRP and the LO phase are chosen at random and independently by the verifier, and they are never revealed.

\begin{figure}[t]
\centering
\includegraphics[width=0.6\linewidth]{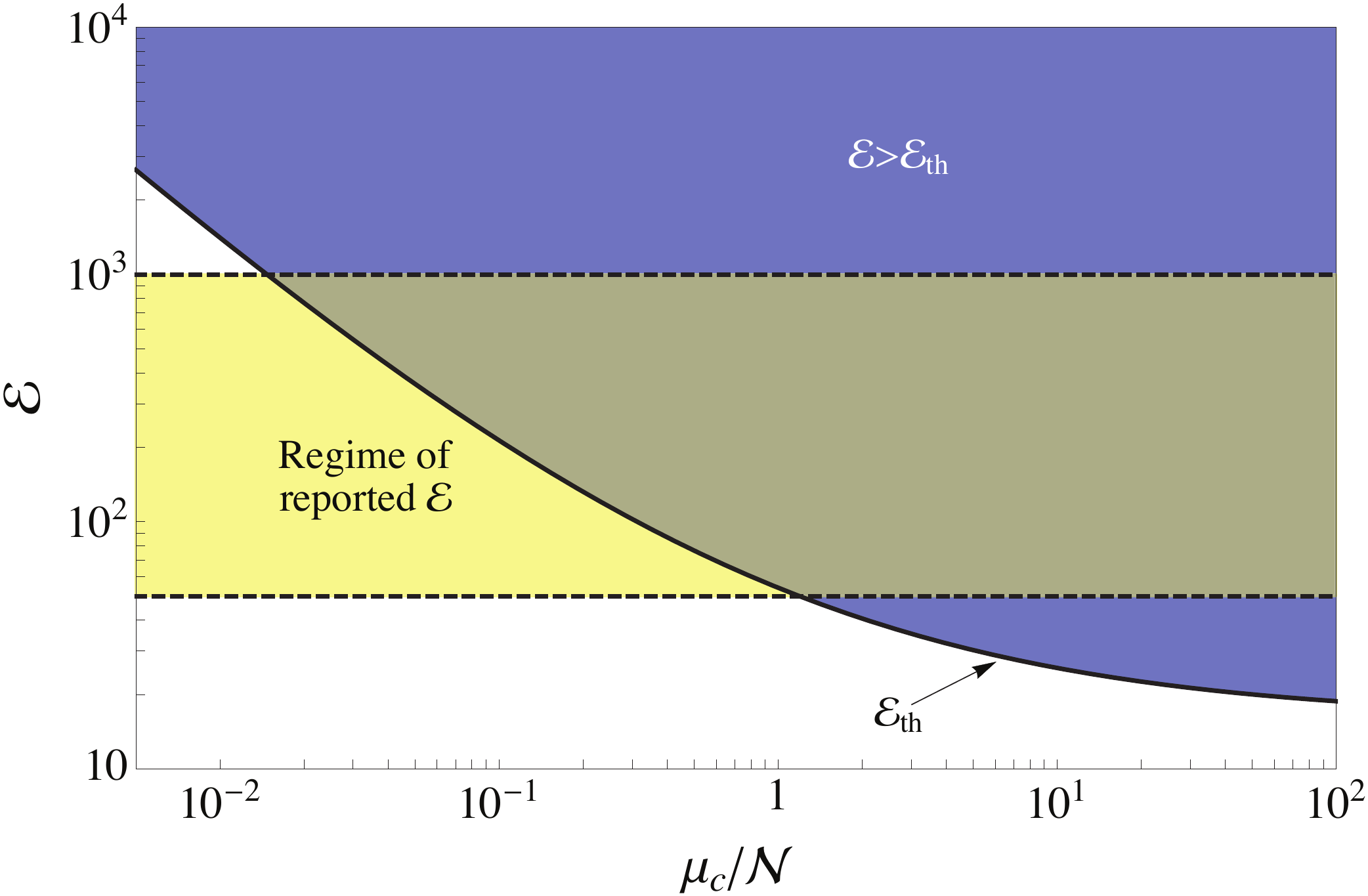}
\caption{
Condition (\ref{Eth:eq}) for various values of the mean photon number per incoming mode. The horizontal band marks the regime of reported enhancements in the literature for different wavefront-shaping set-ups, which range from 50 to about 1000 \cite{Yilmaz13}.  Parameter $l/L = 0.2$.
}
\label{fig:enh}
\end{figure}


\subsubsection{Collision resistance}

Collision resistance refers to the protocol's capability of distinguishing between two randomly chosen keys, and its importance is twofold \cite{handbook,handbook2,Pappu02,PappuPhD}. First, it implies that the EAP can distinguish between different honest users who are holders of random and independently
prepared keys. Second, it is not possible to cheat on a collision-resistant EAP by using a randomly chosen false key.

To gain some insight into the operation of the EAP, let us focus first on a single session, with the typical situation for the response of the true key and a false key summarized in Fig.~\ref{fig2}. The main observation is that, with high probability, the response of a false key lies close to the origin $(0,0)$ of the phase representation shown in Fig.~\ref{fig2}(a), inside or very close to a  circular area of radius  $\rho_f=4\sqrt{\mu_c{\cal V}}$ [see dashed circle in Fig. \ref{fig2}(a)], whereas the response of the true key lies well outside this area [see star in Fig. \ref{fig2}(a)], with its precise location determined by the enhancement ${\cal E}$ and the probe state.
This behaviour can be explained easily, if we note that the phase mask of the SLM is not optimized with respect to the false key, and Eqs. (\ref{XY:nonopt}) imply that
the quadratures of the scattered field will satisfy $|\aver{\hat{X}_s}_{\rm no}|\lesssim \rho_f$ and $|\aver{\hat{Y}_s}_{\rm no}|\lesssim \rho_f$, with high probability.
By contrast, when the true key is interrogated, the SLM is optimized, and from Eq.~(\ref{optXY:condition}) we have that either $|\aver{\hat{X}_s}_{\rm o}|\geq \rho_t$ or $|\aver{\hat{Y}_s}_{\rm o}|\geq \rho_t$, with $\rho_t :=\sqrt{{\cal E}}\rho_f/4$. For the parameters used in Fig.~\ref{fig2}(a), we have $\rho_f = 10$, ${\cal E} \simeq \pi {\cal N}/4 \simeq 201$, and $\rho_t\simeq 35.5$, which correspond to the depicted behaviour.

These observations hold for any session, where in each session the verifier chooses at random the quadrature to be measured, and checks whether the outcome falls within a bin that is centred at $\aver{\hat{X}_s}_{\rm o}$ or $\aver{\hat{Y}_s}_{\rm o}$, for $\theta=0$ and $\pi/2$ respectively.
Given that both quadratures $\hat{X}_s$ and $\hat{Y}_s$ are treated equally, after $M\gg 1$ sessions the verifier has obtained samples from both distributions.
As discussed earlier, for the true key the centres of the sampled distributions coincide with the centres of the bins and, irrespective of the measured quadrature, the theoretical probability for the outcome to fall in the bin is $P_{\rm in}$ [see Fig.~\ref{fig2}(b)]. By contrast, in the case of a false key, the samples are obtained from Gaussian distributions of the form of Eq.~(\ref{PrHD:eq}), centred at $\aver{\hat{X}_s}_{\rm no}$ and $\aver{\hat{Y}_s}_{\rm no}$, and we have either
$|\aver{\hat{X}_s}_{\rm o}|-|\aver{\hat{X}_s}_{\rm no}|\gtrsim \rho_t - \rho_f$ or $|\aver{\hat{Y}_s}_{\rm o}|-|\aver{\hat{Y}_s}_{\rm no}|\gtrsim \rho_t - \rho_f$.
Hence, recalling that $\rho_t = \sqrt{\cal E} \rho_f/4$, and for sufficiently large values of ${\cal E}$, we expect negligible overlap between one of the sampled distributions and the corresponding bin [see Fig.~\ref{fig2}(b)], thereby resulting in a significant deviation of the empirical probability $p_{\rm in}$ from the theoretical probability $P_{\rm in}$.

\begin{figure}[tb]
\centering
\includegraphics[width=0.8\linewidth]{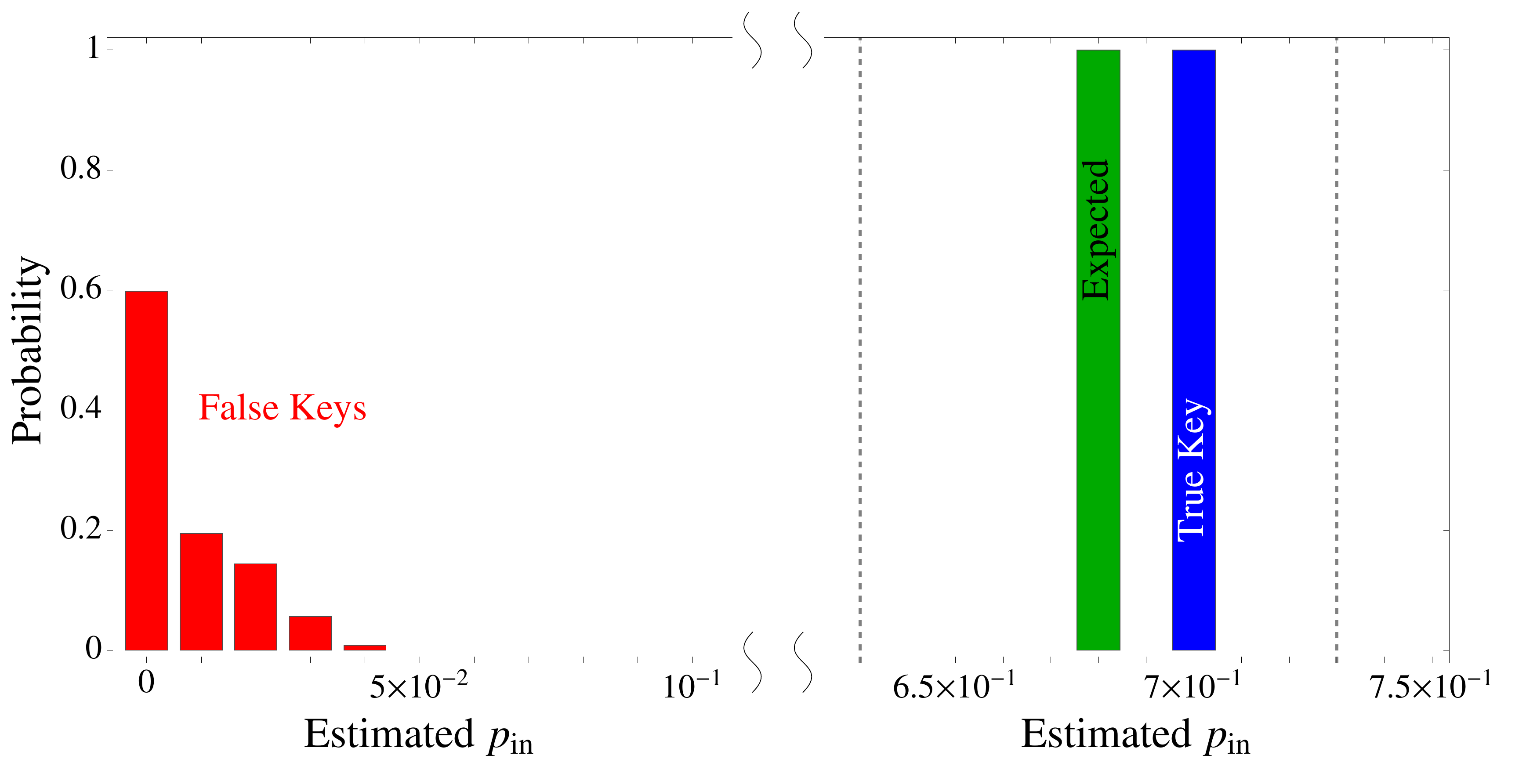}
\caption{
Monte Carlo simulation of the protocol with $M = 10^3$ sessions.
Each red bar gives the probability for a false key to result in an estimate $p_{\rm in}$ that lies in an interval $[p, p+dp)$. We also show the theoretically expected probability $P_{\rm in}$, given by Eq. (\ref{Pexp:eq}), together with the estimate for the true key for the given $M$. The vertical dashed lines define $P_{\rm in}\pm\varepsilon$. The probabilities have been obtained by simulating the verification of 500 randomly chosen false  keys, as well as the verification of the true key (for which the phase mask of the  SLM is optimized).
Note that the histogram for the false keys is peaked at a distance which is about an order of magnitude away from $P_{\rm in}$, whereas the estimate for the true key satisfies $|p_{\rm in} - P_{\rm in}|<\varepsilon$.
Parameters: ${\cal N} = 121$ modes, uniform illumination of SLM, $\tau=0.8$,  $\eta=0.55$, $\delta=2\sigma$, $dp=0.01$, $\varepsilon = 0.05$, $l/L=0.2$, $\mu_P = 2500$, $N=11$.
}
\label{fig4}
\end{figure}

In practice, we would like to have as large deviations as possible so that the detection of a false key is guaranteed. To this end, it is sufficient to impose the condition $\rho_t-\rho_f\gtrsim 3\sigma$. This is because, according to the Gaussian distribution of Eq.~(\ref{PrHD:eq}), outcomes $q$ with $|q-\aver{\hat{Q}_s(\theta)}|\gtrsim 3\sigma$ occur with probabilities that are at least two orders of magnitude smaller than the maximum probability corresponding to the outcome $q=\aver{\hat{Q}_s(\theta)}$. Assuming $0.5\leq \eta \leq 1$, the worst case scenario is for $\sigma = 1$, and using the above expressions for $\rho_t$ and $\rho_f$ one readily obtains ${\cal E}\gtrsim 16(1+3\rho_f^{-1})^2$ or else
\bea
{\cal E}\gtrsim 16\left \{1+\frac{3}{4}\left [ \frac{\mu_{\rm c}}{\cal N} \left (1-\frac{l}{L} \right )\right ]^{-1/2}\right \}^2:= {\cal E}_{\rm th}.
\label{Eth:eq}
\eea
Condition (\ref{Eth:eq}) ensures the detection of a false key, because it implies that for at least one of  the quadratures, the corresponding distribution has negligible overlap with the bin used by the verifier, and hence that it will have negligible contribution to the estimation of $p_{\rm in}$.
Typically, the number of modes ${\cal N}$ and the enhancement ${\cal E}$ depend strongly on the details of the wavefront-shaping set-up, whereas the fraction $l/L$ depends only on the key. For a fixed wavefront-shaping set-up, and assuming that the keys used in the EAP are characterized by the same ratio $l/L$, one can easily adjust the mean number of photons per incoming mode, $\mu_{\rm c}/{\cal N}$, so that the above condition is satisfied.
As shown in Fig. \ref{fig:enh}, condition (\ref{Eth:eq}) is satisfied in many existing wavefront-shaping set-ups, for a broad range of mean photon number per mode values.

To confirm the above observations, we have performed simulations of the EAP for various combinations of parameters. More details about our simulations can be found in the Methods section, and in Fig. \ref{fig4} we present an example of our results. Clearly, with high probability the false key results in an estimate $p_{\rm in}$, which is about an order of magnitude smaller than the expected probability $P_{\rm in}$, and thus it will be detected by a verification test with any error $\varepsilon<1$.
At the same time the true key results in an estimate that satisfies $|p_{\rm in} - P_{\rm in}|<\varepsilon$, and thus it will pass the verification test.
Condition (\ref{Eth:eq}) is readily satisfied for the parameters used in Fig.~\ref{fig4} (we have ${\cal E} \simeq \pi {\cal N}/4 \simeq 95$ and ${\cal E}_{\rm th} \simeq 23$), leading to the depicted difference between $P_{\rm in}$ and $p_{\rm in}$ in the case of a false key.

\begin{figure}[tb]
\centering
\includegraphics[width=1\linewidth]{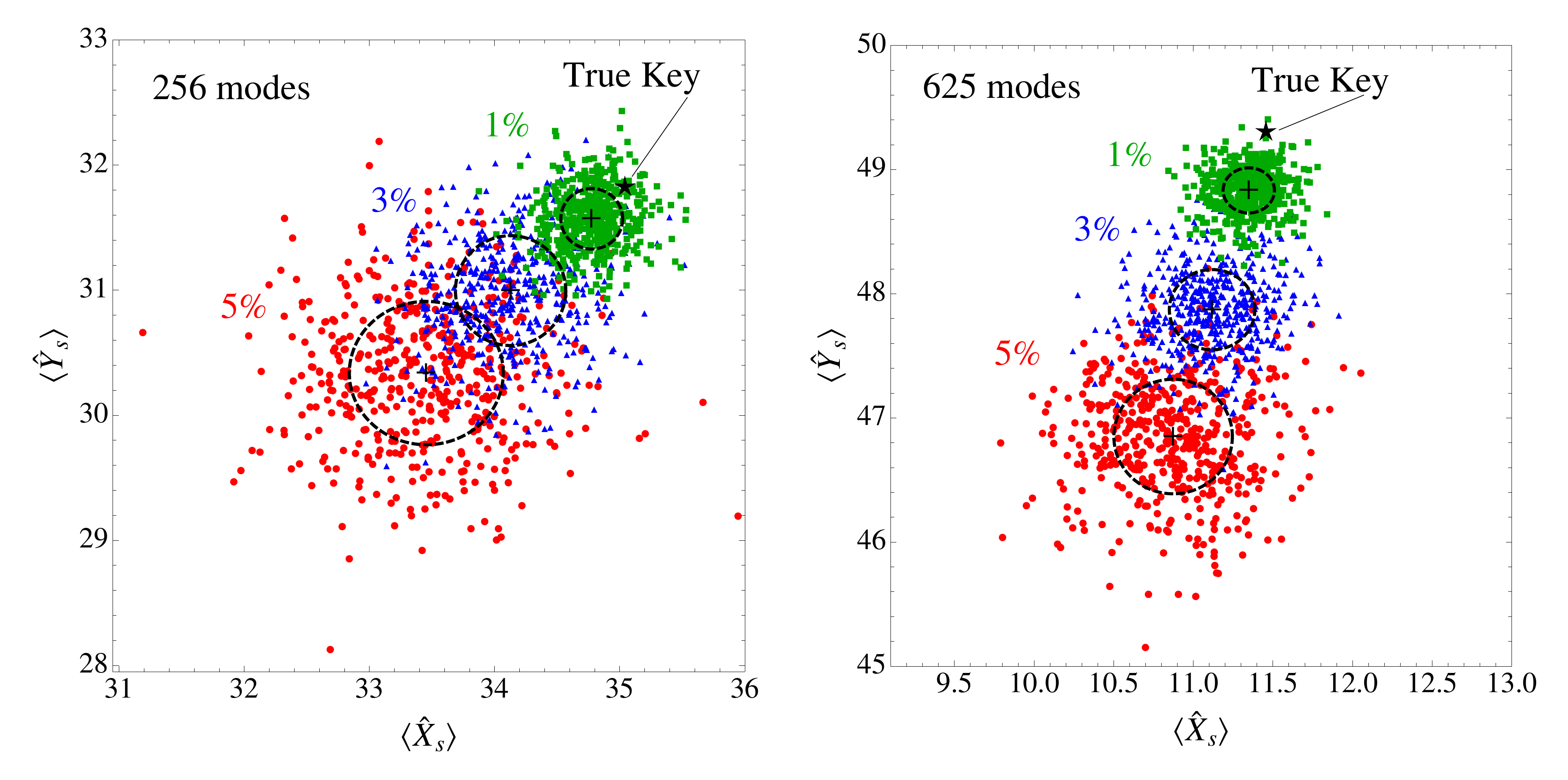}
\caption{
Typical response of $D-$close clones (coloured symbols) relative to the response of the true key (star). The responses of 500 random $D-$close clones is shown in phase-space representation for various values of $D$ (1-5\%), and two different numbers of modes. The responses of the clones  move away from the response of the true key, as we increase $D$. Also shown are the mean value (black cross) and the standard deviation (dashed circle) of the responses of the $500$ random $D$-close clones, for each $D$. Parameters are as in Fig. \ref{fig4}.
}
\label{fig5}
\end{figure}


\subsubsection{Sensitivity to cloning}

Although perfect cloning of PUKs is considered to be practically impossible, imperfect cloning cannot be excluded \cite{Pappu02,PappuPhD,Goorden14}. The question therefore is whether our EAP is capable of distinguishing between the true key and a clone of it.
To address this question, we modelled a $D-$close clone by a scattering matrix, which differs from the scattering matrix of the true key in a fraction of elements $ D\leq 1$. Hence, the quality of the clone increases with a decreasing $D$, with $D = 0$ and $D=1$ corresponding to a perfect and a totally randomized clone, respectively. A $D-$close clone is expected to pass the verification test if its response to the $M$ random challenges is such that, with high probability, the estimated probability $p_{\rm in}$ satisfies $|p_{\rm in} - P_{\rm in}|<\varepsilon$. But, how good a clone should be in order for this to happen?

\begin{figure}[t]
\centering
\includegraphics[width=0.7\linewidth]{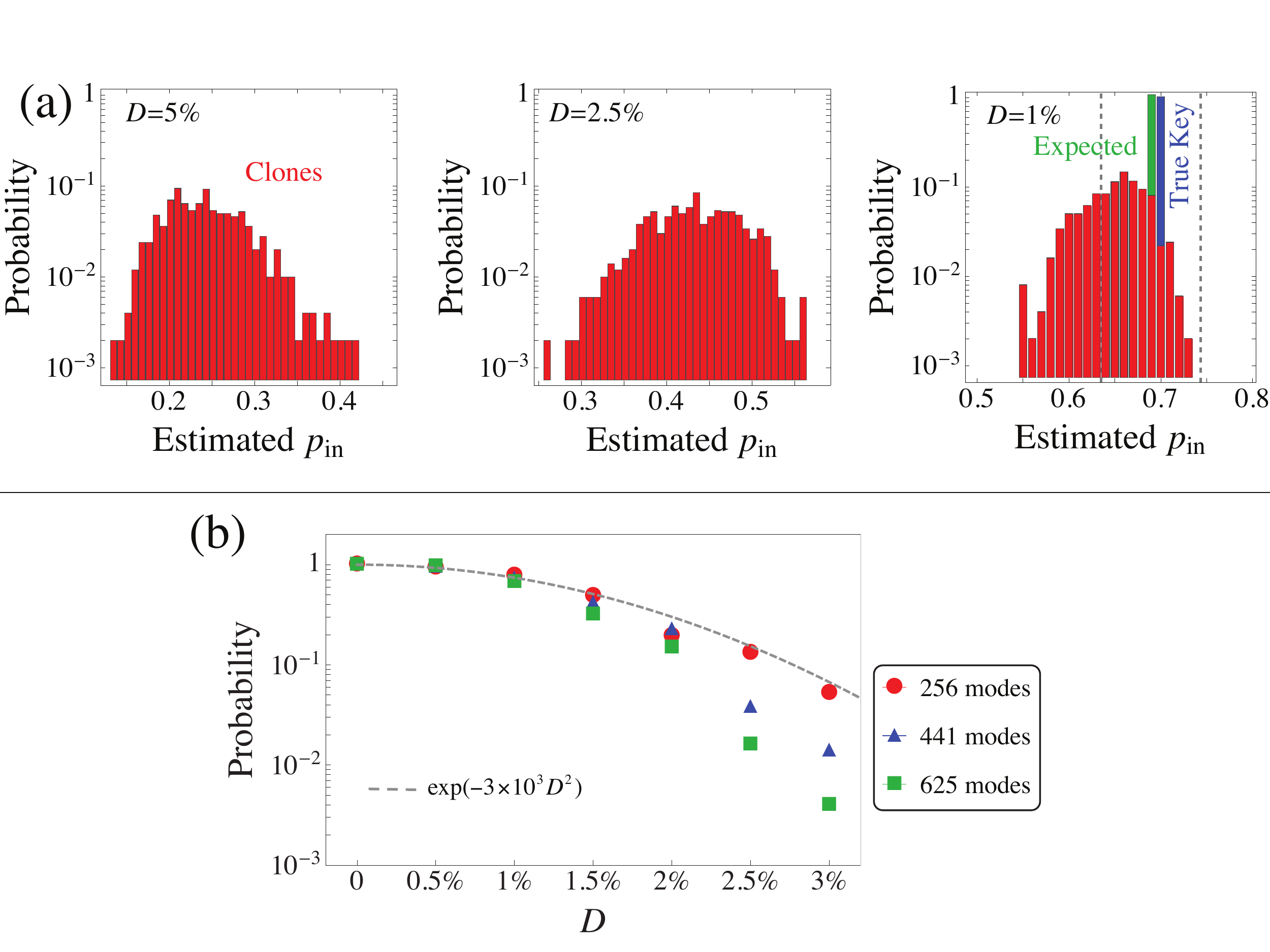}
\caption{
Monte Carlo simulation of the protocol with $M = 10^3$ sessions.
(a) The probability for a $D-$close clone of the true key  to result in an estimate $p_{\rm in}$ that lies in an  interval $[p, p+dp)$. The theoretically expected probability $P_{\rm in}$,  together with the estimate for the true key, are also shown on the right histogram. Note the logarithmic scale of the vertical axes and the different scales in the horizontal axes. The histograms are shown for ${\cal N}=625$ modes, but analogous behaviour has been found for all of the values of ${\cal N}$ we have studied. It is only for very small values of $D$ that there is a significant probability for a clone to yield $p_{\rm in}$ close to $P_{\rm in}$ (see histogram on the right).
(b) Probability of successful cheating, i.e., the probability for a $D-$close clone to escape detection in a verification stage with $\varepsilon = 0.05$, for three different numbers of modes. The fitting curve (dashed line) has been plotted to guide the eye.
For both (a) and (b), the probabilities for a given $D$ have been obtained by simulating the verification of 500 independent randomly chosen $D-$close clones. Other parameters are as in Fig. \ref{fig4}.
}
\label{fig6}
\end{figure}

The typical response of $D-$close clones relative to the response of the true key is shown in Fig. \ref{fig5}. We see that for values of $D\lesssim 1\%$, the response of $D-$close clones lies very close to the response of the true key. In this case, one may expect high probability for a clone to result in a probability $p_{\rm in}$ very close to $P_{\rm in}$. As $D$ increases, however, the responses of the clones move rapidly away from the response of the true key, and $p_{\rm in}$ is also expected to move away from $P_{\rm in}$. This behaviour is clearly shown in the probability distributions of Fig. \ref{fig6}(a). As a result, the probability for a $D-$close clone to pass the verification test, i.e., to result in an estimate $p_{\rm in}$ such that $|p_{\rm in} - P_{\rm in}|<\varepsilon$, decreases rapidly with increasing $D$ [see Fig. \ref{fig6}(b)]. Note that for fixed ${D}$ and ${\cal N}$, this probability is expected to decrease with decreasing error $\varepsilon$, because the accuracy in the estimation of $P_{\rm in}$ increases in this case.
Figure \ref{fig6}(b) suggests that for ${\cal N}\geq 256$ and $\varepsilon \leq 5\times 10^{-2}$, 
the  scattering matrix of a clone should differ from the one of the key in a small fraction of elements (smaller than $3\%$ or so), in order for the clone to have a non-negligible probability to pass the verification test. Cloning of such a high quality is a formidable challenge for today's technology, because it requires the exact positioning (on a nanometer scale) of millions of scatterers with the exact characteristics \cite{Pappu02,Goorden14}. It is also worth noting here that according to the results of Figs. \ref{fig5} and \ref{fig6}(b), the robustness of the EAP  against cloning appears to increase considerably with an increasing number of modes. This finding suggests that if the protocol is realized using existing  wavefront shaping set-ups, which have been shown capable of controlling  thousands of modes,  then the probability for a clone with $D\approx 3\%$ to pass the verification test will be at most $10^{-3}$.

We remark that we have performed simulations for many different combinations of parameters, but for practical reasons we have presented results for certain representative combinations only. The main findings and conclusions presented here hold for all of the combinations we have studied, and we expect that they are generally valid.

\section{Discussion}

In the present form of the EAP, the number of sessions that can be performed within a prescribed period of time, is mainly limited by the separation distances of the various components of the set-up, and the HD bandwidth. Assuming that the different components of the set-up (laser source, interrogation chamber and HD set-up) are located in neighbouring rooms, the typical total distances to be travelled by the probe and the scattered light are of the order of tens of meters. HD bandwidth is typically $\sim 10 - 100$ MHz depending on the specific implementation. Hence, the time of a single session, i.e., the time that it takes for a pulse to propagate from the laser source to the key, and from there to the HD set-up where it will be analysed, is estimated to be less than a microsecond. According to Eq.~(\ref{M_th:eq}), verification tests of error $\varepsilon\simeq 1\times 10^{-3}$ and confidence $99.9\%$ require $M\simeq 2.3\times 10^7$ sessions, and the total verification time is expected to be a few seconds.  

Our EAP is the first one to rely on conjugate quantum continuous variables, and provides a practical way to secure authentication of optical PUKs without the need for photon counting.
Assuming a tamper-resistant verification set-up, we have shown that the protocol offers collision resistance and robustness against cloning. Moreover, it is worth emphasizing  that, as long as the verification set-up is tamper resistant, a compromised  database does not affect the security of the protocol. Indeed, even if an adversary has access to the list of CRPs to be used for the authentication of a key, the sequence of probe states as well as the sequence of the quadratures to be measured in $M$ sessions, are not {\em a priori} known. They are chosen at random during the verification, and the  probability for an adversary to  guess correctly both sequences is $(2|{\mathbb A}|)^{-M}\ll 1$, for $|{\mathbb A}|,M\gg 1$.

Collision resistance and robustness against cloning are necessary for our EAP to be useful in practice \cite{handbook,handbook2,Pappu02,PappuPhD}. Its security against cheating strategies, where an adversary has access to the verification set-up, goes beyond the scope of the present work, and requires an in-depth description and analysis of the  strategy under consideration. We do point out, however, that a prerequisite for the successful implementation of such attacks is that the adversary has access to the challenge states (or equivalently to the interrogation chamber), as well as to the LO, without being noticed by the verifier. The proposed fiber-based implementation of our scheme allows for the spatial separation of the interrogation chamber from the laser source and the HD set-up (e.g., they may be located in nearby rooms). The LO never enters the interrogation chamber, and an adversary who has access to this chamber only does not have access to the reference frame used for the definition of the quantum state
of the probes. Finally, it is worth emphasizing that the only constraints on the mean number of photons of the probe are the ones imposed by Eq. (\ref{Eth:eq}).
This is because we have assumed that the verification set-up is tamper-resistant. The security 
of the protocol against attackers who have access to the verification set-up may require 
additional constraints on the mean number of photons in the probes, as well as on the size of 
the set of states $|{{\mathbb A}}|$. Such security analysis depends strongly on the details of 
the attack under consideration, 
but it  will rely on the fact that the quadrature components of the electric field do not 
commute,  and thus by virtue of Heisenberg's uncertainty relation,  they cannot be 
determined  simultaneously with arbitrary accuracy.

\appendix

\section{Methods}

\subsection{The Gaussian-statistics model}
The model (\ref{Rstat:eq}) assumes that we are in the diffusive regime, and the key consists of a large number of independent totally unrelated elementary scattering areas\cite{Goodman1,Anderson14}. The electric field of the scattered light at a particular observation point consists of a multitude of de-phased contributions from different scattering areas, and thus its amplitude can be expressed as a sum of many elementary phasor contributions.  As a result of the occurrence of multiple scattering events, the 
amplitude of each phasor bears no relation to its  phase, while the latter is uniformly distributed over $[-\pi,\pi]$. Under these conditions, the central-limit theorem implies that the scattering problem can be described in the framework of a  scattering matrix with independent identically distributed random entries of Gaussian statistics. 
These conditions have been shown to be satisfied in many experimental set-ups \cite{Goodman1,book2,Lodahl_prl05,Lodahl_oe06}, and the Gaussian-statistics model yielded results that were in excellent agreement with experimental observations.  

\subsection{Sample size in the enrolment stage}
The two quadratures of ${\cal R}_s^{({\rm o})}(\alpha_k)$ are the centres of Gaussian 
distributions of standard deviation $\sigma$. 
We assume that the quadratures are estimated by 
sampling at random and independently from the corresponding Gaussian distributions. Consider one of the quadratures, say $\aver{\hat{X}_s}_{\rm o}$. 
It will be approximated by the sample mean, which is also a random variable and  
according to the central-limit theorem, it follows a Gaussian distribution centred 
at $\aver{\hat{X}_s}_{\rm o}$ 
and with standard deviation $\widetilde{\sigma} = \sigma/\sqrt{M_e}$, where $M_e$ is the sample size. 
Hence, the probability to obtain estimates outside the interval 
$[\aver{\hat{X}_s}_{\rm o}-5\widetilde{\sigma},\, \aver{\hat{X}_s}_{\rm o}+5\widetilde{\sigma}]$ 
is at most $10^{-6}$.  
In other words, it is highly unlikely for the error in the estimation of the quadrature to exceed 
$\xi=5/\sqrt{M_e}$, where we have assumed that $\sigma\leq 1$. 
As mentioned in the main text, the sample size  $M_e$ has to be such 
that $\xi\ll \varepsilon$, where $\varepsilon$ is the error in the verification test. 
These arguments hold for the estimation of either of the 
two quadratures for a given probe state, and assuming that both quadratures and all of 
the probe states are treated equally, the total sample size is 
$M_e^{{\rm (t)}} = 2\times N\times M_e$ ( for a standard HD set-up). 
Even for moderate values of $N>2$, the total sample size $M_e^{{\rm (t)}}$ is expected to 
be considerably larger than the sample sizes typically used in the verification stage. 
Assuming  identical enrolment and verification set-ups, we can use the parameters 
of the Discussion above, in order to obtain an estimate for the duration of an enrolment 
stage with $N=10$, $\varepsilon\simeq 10^{-3}$ and $\xi \simeq 0.1\varepsilon$.  
The total sample size is $M_e^{{\rm (t)}} \simeq 5\times 10^{10}$, while the typical sample time is expected to be less than a microsecond. 
Hence, the enrolment stage will last less than 14 hours, which is not in any case prohibitive, given that the enrolment is performed only once by the manufacturer, well before a key is given to a user.
 
\subsection{Sample size in the verification stage} 
In our EAP, the verification relies on the estimation of the probability $P_{\rm in}$, which 
refers to the probability for an outcome that is drawn at random from the Gaussian distribution 
(\ref{PrHD:eq}), to fall within an interval (bin) of size $\delta$. 
To estimate the sample size (i.e., the number of sessions) required for the reliable estimation of 
$P_{\rm in}$, 
we can introduce  a binary random variable for the $i$th session, say $\omega_i$, which
refers to whether the outcome of the measurement in the $i$th session falls or not within the 
specified bin. 
In particular, let $\omega_i$ be $1$ when the outcome lies inside the interval, and $0$ otherwise. 
The former occurs with probability $P_{\rm in}$, and the latter with probability 
$P_{\rm out}:=1-P_{\rm in}$.
Recall that all of the sessions in the verification are identical, and independent of each other.
For $M$ sessions, we can introduce the random variable $M_{\rm in}:= \sum_i \omega_i$, and let 
$p_{\rm in}:=M_{\rm in}/M$ be an estimate of $P_{\rm in}$ based on the outcomes in $M$ sessions.  
Our task is to estimate how large $M$ must be in order for the estimate 
to satisfy ${\rm Pr}[|p_{\rm in}-P_{\rm in}|< \varepsilon] > 1-\zeta$, 
where $\zeta \ll 1$ is the uncertainty, and $\varepsilon\ll P_{\rm in}$ the absolute error. 
To this end, it is sufficient to ask for 
\bea
{\rm Pr}[|p_{\rm in}-P_{\rm in}|<\tilde{\varepsilon} P_{\rm in}] > 1-\zeta, 
\label{rel_err:ChB}
\eea
where $\tilde {\varepsilon}:= \varepsilon/P_{\rm in}\ll 1$. 
From the Chernoff's bound \cite{book4,book5} for the relative error we have 
\bea
{\rm Pr}[|p_{\rm in}-P_{\rm in}|\geq \tilde{\varepsilon} P_{\rm in}] \leq 2\exp \left (
-\frac{\varepsilon^2 M}{3}
\right ), 
\eea
where we have used the inequality 
$\frac{\varepsilon^2 M}{3 P_{\rm in}}> \frac{\varepsilon^2 M}{3}$ for $0<P_{\rm in}<1$. 
To enforce condition (\ref{rel_err:ChB}), we ask for the upper bound in the last expression to  
be less than $\zeta$. Subsequently, solving for $M$ one readily obtains that the sample size has to be 
larger than $M_{\rm th}$, where $M_{\rm th}$ is given by Eq. (\ref{M_th:eq}). 
 
\subsection{Simulations}
We performed simulations for various combinations of parameters, and for each set of parameters we worked as follows. We generated ${\cal N}$ reflection coefficients of the true key, using a generator of complex Gaussian random variables with the characteristics of Eqs. (\ref{Rstat:eq}). Subsequently,  we found the optimal phase mask of the SLM that maximizes the number of scattered photons in a prescribed target mode, using known algorithms \cite{Vellekoop08}. The false keys and the clones were generated along the same lines. Each false key pertained to a set of ${\cal N}$ random and independently chosen reflection coefficients, whereas  for a $D-$close clone we substituted $D\times{\cal N}$ of the reflection coefficients of the true key by fresh random and independently chosen coefficients. The elements that were substituted were also chosen at random and independently.

Each key (true, false or clone) was interrogated by $M$ probes, with the state of each probe chosen at random and independently from a uniform distribution over a finite set of prescribed coherent states.
In each session, i.e., for each probe state, $\aver{\hat{b}_{s}}$ was obtained from Eq. (\ref{avBnonuni}), using the reflection coefficients for the true key, the false key, or the clone, while the phase mask for the SLM was always set to the optimal configuration that maximizes the light that is scattered from the true key to the target mode $s$.  Having estimated  $\aver{\hat{b}_s}$, we chose $\theta$ at random from a uniform distribution over $\{0,\pi/2\}$. In accordance with the theory of HD, the outcome of a measurement of the quadrature $\hat{Q}_s(\theta)$ was simulated by a real random variable, which was chosen from a Gaussian distribution centred at $\aver{\hat{Q}_s(\theta)}$ and with standard deviation $\sigma=1/\sqrt{2\eta}$.
At the end of the session we checked whether the outcome falls within the bin ${\mathfrak B}[{\cal R}_s^{({\rm o})},\theta]$ or not.

By performing this procedure for a large number of random and independently chosen false keys, and clones, we obtained sufficiently large samples to estimate the probabilities shown in the figures. It is worth emphasizing that different random generators were employed in our simulations, so that to ensure independence of the drawn random numbers. Finally, we note that for practical and numerical reasons, the number of sessions in our simulations could not exceed $10^3$. Our results, however, show that this number was sufficient for the verification stage to identify successfully the true key and to detect the false keys and the clones, which suggests that $M_{\rm th}$ is not a tight lower bound on the required number of sessions.

\section*{Acknowledgements}

This article is based on work performed in the context of the Nanoscale Quantum Optics COST Action (MP1403), supported by the European Cooperation in Science and Technology. We acknowledge financial support from the French National Research Agency project QRYPTOS.

\section*{Author contributions statement}

GMN  conceived the main idea, developed the theory and  performed the simulations. ED contributed to the analysis of the results, as well as to practical aspects pertaining to the implementation of the protocol.

\section*{Additional information}

The authors declare no competing financial interests.

\end{document}